# A Numerical Investigation of Three-dimensional Falling Liquid Films

**Idris Adebayo**[1,2]

*Department of Chemical Engineering, Imperial College London, SW7 2AZ, UK*
*Department of Chemical Engineering, Federal University of Technology, Minna, Nigeria*

Email: *Idris.adebayo13@alumni.imperial.ac.uk; idris.adebayo@futminna.edu.ng*
ORCID iD: 0000-0002-1682-9628

**Abstract** The flow of thin liquid films on inclined or vertical surfaces is one of immense importance, with applications spanning many types of process industries, due to the increased mass and heat transfer brought about by the presence of waves on film surfaces. While extensive research has been carried out in this area, many outstanding questions remain due to the complexity of the problem - a result of the extremely large aspect ratio and the three-dimensional nature. In this study, the evolution process of both naturally forming and forced waves on these thin liquid films is numerically studied using a hybrid front-tracking/level-set hybrid method which has the advantage of a distinct and accurate capture of the interface topology without being affected by the issues of mass conservation, numerical diffusion or spurious interface profiles. The simulation is conducted on an "infinite" domain specification through the imposition of periodic boundary conditions at the film outlet. The obtained result show the growth, step-wise progression and eventual transition of natural waves from the two-dimensional state to three-dimensional disordered structures. Finally, three-dimensional simulations of artificially forced waves are carried out while different parameters/features of these waves are compared with the existing two-dimensional benchmark cases in the literature with excellent agreement found.

# 1 Introduction

The flow of thin liquid films constitutes a very important research area in fluid dynamics. This owing in parts to the interesting dynamics of wave evolution observed on these films as well as the numerous application it has in many types of process equipment [1,2]. These applications which are seen in cooling, condensation, evaporation, adsorption processes, etc. also come with the advantage of intensified heat and mass transfer rates provided by the wave structures on these film, even at relatively low liquid flow rates and short contact time between the liquid and solid wall [3-6].

An extensive effort has been put into its research following the initiating work of father and son, Kapitza and Kapitza [7], who studied the evolution of waves on thin films, with Re number range between 7-23. Gollub and Co-workers [8,9] also carried out extensive research into both developments of large-amplitude solitary waves (otherwise referred to as gamma II waves) and the secondary instabilities following the initial bifurcation from a flat film steady state. Their results revealed the evolution of structured patterns i.e. "herringbone" from two-dimensional waves due to a Subharmonic instability, and a synchronous three-dimensional mode arising from a sideband instability. Park and Nosoko [10] also performed experiments centered on the artificial excitation of three-dimensional wave structures on the film by the introduction of a transversal disturbance (i.e. a needle) put in contact with the film's free surface. And they obtained a critical perturbation wavelength for the solitary wave instability.

Nosoko et al. [11] also investigated the characteristics of waves on thin films while deriving correlations for the dependence of both wave speed and peak height on the wavelength as well as film Re and We numbers. Further research efforts include the works of Kofman *et al*, [12], Nosoko and Miyara [13], Dietze *et al.* [14] while detailed reviews of several of these studies have also been carried out [1,15] as well as many textbooks authored in this field [2,15-16].

On the mathematical modeling work, initial work was first carried out by Yir [17-18] and Benjamin [19] on the linear stability analyses of the base state of thin films. Their results showed the base state of the film to be unstable to long-wavelength disturbances. Consequently, further studies also focussed on the development of simplified equations from the full Navier-Stokes by a long-wave approximation and with different order-of-magnitude assumptions for the important *Re* and *We* numbers [20-21]. While the wave behaviour has still not been fully understood, since these theoretical studies are only possible in the framework of linear stability analysis, which is limited to the initial stage of wave inception and very small Re values; efforts have also been made on the solution of the non-linear equations. However, the analytic solutions are always limited to trivial cases. Hence, the focus on numerical solutions.

With numerical simulations, a number of researches have also been carried out. These include Direct Numerical Simulation by Miyara [22], using a simplified marker and cell method. Ramaswamy *et al.* [23] using Finite Element Method with Eulerian-Lagrangian formulation, D. Gao [24] on falling films on vertical planes using the Volume of Fluids method with a Continuum Surface Force for surface tension, Yu *et al.* [25], Xie *et al.* [26] and recently Adebayo *et al.* [27] with an algebraic formulation of the Volume of Fluids using unstructured, anisotropic and adaptive meshes.

However, many of the above-mentioned works have only considered the two-dimensional analysis of the falling film problem. Though this two-dimensional solution seems to provide a wide range of information as to the dynamics of the wave evolution process, it must be noted that the interface topology usually exhibits serious transversal changes which are impossible to observe in the two-dimensional plane, hence leaving a far more abundant amount of information to be discovered in the three-dimensional analysis. However, though this three-dimensional simulations of the falling film is indisputably superior to the two-dimensional analysis, the complexity of the problem and the computational requirement to resolve it has been one of the caveats to its study. In this study, we present results on a full three-dimensional direct numerical simulation of this thin-film flows.

As a background, it must be understood that while a number of approaches have been successfully applied to resolving the fluid properties at the bulk (*e.g.* Finite Element, Finite Volume, Finite-difference, etc.), the presence of an interface in two-phase flows poses a challenge, considering the need to accurately determined the interface shape and location in time and also trail the fluid properties which may change abruptly due to discontinuities at this interface. These methods for interface treatment have been classified into two (1) Front capturing: Volume of Fluid, Level set, and phase-field which represents the interface implicitly on a fixed Eulerian mesh and (2) Front tracking which tracks the interface by using a separate lagrangian discretization of the interface. These Front capturing methods are based on an extra field variable (i.e. the volume fraction function) which takes a value of 0 or 1 in the bulk of a fluid while cells with values between 0 and 1 exclusive denote the interface presence. In practice, the front capturing methods are generally simpler since they don't require any additional grid but only advect the interface using high order schemes (adopted from high-speed compressible flow). The Front tracking method, however, shows the best performance in maintaining the sharpness of the interface as the lagrangian marker points defining the interface are only integrated in time. Unfortunately, the disadvantage of the front tracking method is observed in the difficulties involved in the separate grid depended upon to accurately model the complex phenomena occurring at the interface.

While most parallel two-phase codes simple use the front capturing methods due to the ease of parallelizing the Eulerian field data [28], the Advantages of front tracking: accurate mass conservation; no numerical diffusion; accurate sub-grid description of interfacial physics; no need for highly refined grids; accurate computation of surface tension force and other interface sources seem very imminent. The only disadvantage of front tracking is that the interface needs a separate mesh discretized by moving triangular elements which adds complexity to the parallelization process. While the main problem with the original front tracking method was the need to logically connect the interface element and bookkeep changes in the connectivity during element addition, deletion or reconnection as the interface topology changes. This is worsened in three-dimensions. Hence, making parallelization a daunting task as the interface solver becomes difficult to parallelize.

In this work, a hybrid front tracking/level set numerical simulation code is utilised which not only captures the details of the interface accurately but also exempts issues of mass conservation and can be run on a massive number of cores/threads - a major distinguishing factor from many front-tracking approaches. Details on the code are outside the scope of this present work, however, an overview of it is presented in Sec III. The interested reader is referred to Refs. [29-31] for further information on the code.

The rest of the paper is arranged as follows: An overview of the simulation code is presented in Sec 2, while the numerical procedure followed is given in Sec 3 and Set-up in 4. Discussion of the obtained simulation results with a comparison with experimental data is given in Sec 5 while concluding remarks and recommendations for future work are stated in Sec 6.

# 2 Numerical code

The numerical solver used in this work is based on a hybrid front-tracking/level set approach otherwise referred to as the Level Contour Reconstruction Method for interface tacking [29-32] and is capable of being run on a variety of computer architectures; from laptops to supercomputers and on threads of the order of $10^5$. It is adaptable to the simulation of flow problems which include but not limited to flow interactions for immersed solid objects, contact line dynamics, species and thermal transport with phase change, etc. Its implementation ensures first that with a front-tracking approach, the interface shape and position is explicitly and accurately determined even in the presence of highly deforming interfaces while with a Level set; the advantage of a distance function approach for interface reconstruction which automatically and naturally handles breakup and coalescence is harnessed. It is this distance function approach that the code uses; which ensures a computed distance function is utilized for implicit periodic reconstruction of the interface elements without the logical connectivity between the interface elements as was necessary in the original Front Tracking method [31]. However, here, the distance function does not play any role in the interface advection: as it mainly does in the original Level Set approach. No advection equation is ever solved for it. Hence there is no need to preserve mass as always done in Level Set initialization.

It further uses a domain decomposition strategy for parallelization with MPI (Message Passing Interface) wherein all processes are carried out locally in each subdomain (assigned a thread/core). Essentially, the solution developed is that the operations are performed on each triangular interface element separately, independent of other elements. The picture is this: contour lines are made such that neighbouring elements share the same node endpoint (in two-dimensions) and lines (in three-dimension), hence forming implicitly connected surface elements. Field variable data exchange occurs across neighbouring subdomains through a boundary buffer zone. But since the lagrangian interface grids move here, an extended interface concept is used in which additional buffer cells (different from the boundary buffer zone used for velocity boundary conditions etc.) are employed to store and exchange interface data, so that independent subdomain operations can be carried out. This approach of "local operation to other elements" in LCRM hence makes it very easy to carry out on distributed processors [31-33]. A Parallel GMRES and Multigrid solvers for the fluid velocity and pressure solution are further employed which are of importance in the high density and viscosities discontinuities: *e.g.* air and dense oil combinations.

The code structure consists basically of two modules: One for the solution of the incompressible Navier Stokes while the second to solve for the interface using: tracking the phase front, initialization, and reconstruction of the interface when required. The code parallelization is based on the algebraic domain decomposition technique and communication is ensured by data exchange across adjacent subdomains (each a core/thread) by Message Passing Interface (MPI) protocol. The Navier Stokes solver computes the primary variables of velocity and pressure on a fixed uniform Eulerian mesh using the Chorin's projection method [34].

Depending on the nature of the physical problem, numerical stability requirements or the user's preference; one of explicit or implicit time integration can be used and also to either first or second order. Spatial discretization is done using the staggered mesh MAC method [35]. Pressure and distance functions are located at cell centres while velocity components are located at cell faces. All spatial derivatives are approximated by standard second-order centred differences. Finally, the Multigrid Interative method is used for solving the elliptic pressure Poisson equation while a modified Multigrid procedure for distributed processors is employed for the non-separable Poisson equation when a large density ratio two-phase problem is been solved up to $10^5$ density ratios.

## 3 Numerical procedure

The single-field formulation of the Navier Stokes equations for an incompressible Newtonian fluid under isothermal condition is used:

$$\nabla \cdot \boldsymbol{u} = 0 \tag{1}$$

$$\rho\left(\frac{\partial \boldsymbol{u}}{\partial t} + \boldsymbol{u} \cdot \nabla \boldsymbol{u}\right) = -\nabla P + \rho g + \nabla \cdot \mu(\nabla \boldsymbol{u} + \nabla \boldsymbol{u}^T) + \mathbf{F} \tag{2}$$

where **u** is the velocity, $P$ is pressure, **g** is the gravitational acceleration and **F** is the local surface tension force at the interface; described by the hydrid formulation

$$\mathbf{F} = \sigma k_H \nabla I \tag{3}$$

$\sigma$ is the surface tension coefficient assumed constant and $I$ is the indicator function which is zero in one phase and one in the other. It is usually resolved with a sharp but smooth transition across 3 to 4 grid cells and is essentially a numerical Heaviside function. $K_H$ is twice the mean interface curvature field and calculated on the Eulerian grid using

$$\kappa_H = \frac{\mathbf{F}_L \cdot \mathbf{G}}{\sigma \mathbf{G} \cdot \mathbf{G}} \tag{4}$$

where $\mathbf{F}_L$ and G are:

$$\mathbf{F}_L = \int_{\Gamma(t)} \sigma \kappa_f n_f \delta_f(\mathbf{x} - \mathbf{x}_f) ds \qquad \mathbf{G} = \int_{\Gamma(t)} n_f \delta_f(\mathbf{x} - \mathbf{x}_f) ds \tag{5, 6}$$

$x_f$ is a parameterization of the interface, $\Gamma(t)$ while $\delta(\mathbf{x} - \mathbf{x}_f)$ is a Dirac distribution that is non-zero only when $\mathbf{x} = \mathbf{x}_f$. $\mathbf{n}_f$ is the unit normal vector to the interface and $ds$ is the length of

$$\mathbf{V} = \frac{d\mathbf{x}_f}{dt}$$

the interface element.

$$\tag{7}$$

The Lagrangian elements of the interface are advected by integrating with a second order Runge-Kutta method where the interface velocity, V, is interpolated from the Eulerian velocity.

Material properties *e.g.* density, viscosity are defined in the entire domain using the indicator function $I(x.t)$ as:

$$b(\mathbf{x},t) = b_1 + (b_2 - b_1)I(\mathbf{x},t); \qquad (8)$$

where 1 and 2 represents the individual phases and *b* the material property.

## 4 Numerical set-up

The experimental problem was cast as a two-phase (air-water) simulation of a thin liquid film flowing in a cuboidal channel. Due to the thinness of the film and the three-dimensionality of the simulation, the simulation is challenging and could only be realised for a few cases. The details of the numerical set-up, computational domain, initial and boundary conditions as well as obtained results are provided below.

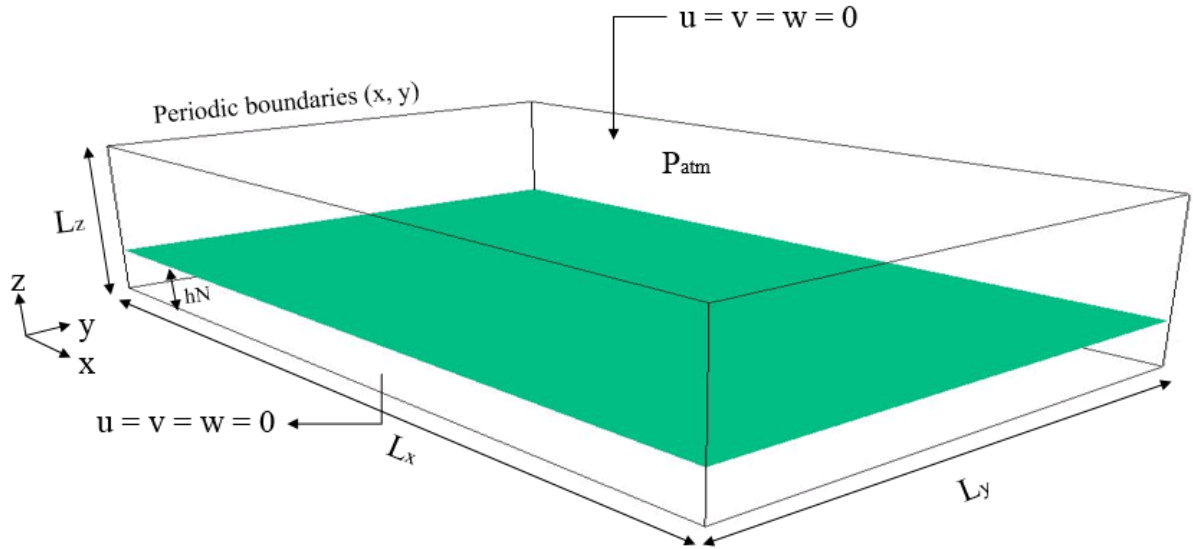

**Fig. 1** The computational domain of the numerical set-up

A three-dimensional simulation of a thin pulsated flowing liquid film was carried out. The density ratio associated with this water-air system was evaluated to be 889 and the simulation has been carried out in a dimensional form on a fixed mesh. Water and air at room temperature with standard properties have been used as respective fluids.

The computational domain is a cuboid represented by $\Omega = (0, x) \times (0, y) \times (0, z)$ (see Fig 1). The x-axis points in the streamwise direction while the y-axis points in the spanwise direction. The position of the interface (separating the gas and the liquid phase) is given by a height function h(t,x,y) over the x and y axes. With this notation, the interface position is determined as:

$$\Sigma(t) = \{(x,y,z) \in \Omega : z = h(t, x, y)\} \quad (9)$$

The domain occupied by the liquid phase is given by

$$\Omega_L(t) = \{(x,y,z) \in \Omega : z < h(t, x, y)\} \quad (10)$$

while the domain occupied by the gas phase is given by

$$\Omega_G(t) = \{(x,y,z) \in \Omega : z > h(t, x, y)\} \quad (11)$$

The dimensions are however; Length 8 cm (in the natural developing case) and 10 cm (in the forced case to allow sufficient wave development). A width of 5 cm and height 0.2 cm is also adopted.

The initial condition consists of a flat film with a parabolic velocity profile determined from the Nusselt theory [36]. At the lower and upper boundaries of the domain, (y=0 and Lz), the no-slip boundary condition is applied while a periodic boundary condition is applied at both the film inlet (x=0) and outlet (x=L).

In the bulk of the film, the initial condition prescribed is given below:

$$U_x(z) = \frac{\rho g \sin\beta}{\mu_w} z \left(h - \frac{z}{2}\right) \quad (12)$$

However, in the case of the forced waves we apply a periodic inlet forcing of the flow rate according to the equations below:

$$U_x(x = 0, \ 0 \leq z \leq h_N) = \frac{3}{2}[1 + A\sin(2\pi F t)]\left(\frac{2z}{h_N} - \frac{z^2}{h_N^2}\right) U_N \quad (13)$$

$$U_x(x = 0, h_N \leq z \leq L_z) = \frac{3}{2}[1 + A\sin(2\pi F t)] U_N \quad (14)$$

Where A is the disturbance magnitude or amplitude (set here to 0.05, as consistent with previous studies [25,27]), and F is the excitation frequency.

The parabolic velocity profile with a temporal periodic disturbance is also consistent with previous numerical studies on excitation of waves on thin liquid films [2-5, 10-14] while Yu *et al.* [25] has shown that the wave development is insensitive to the magnitude of this disturbance amplitude.

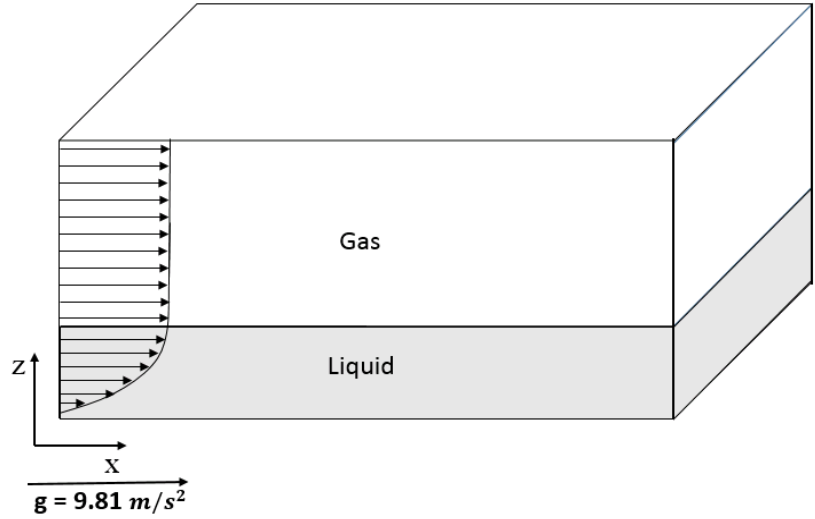

**Fig. 2** Parabolic velocity profile for falling film initialisation.

The simulations have been carried out with water at room temperature with density p=1000kg/m3, kinematic viscosity, v = 1 x10⁻⁶ m²/s, surface tension =73.5 mN/m. The dimensionless groups used include film *Re, Ka, We, Fr*. Details on the ranges of the individual parameters are presented in Table 1.

**Table 1 Table of experimental parameters**

| Variable | Definition | Value/Ranges | Unit |
| --- | --- | --- | --- |
| Substrate inclination angle | $\beta$ | 90 | degree |
| Forcing frequency | F | 15 - 27 | Hz |
| Water viscosity | $\mu_w$ | $1.0 \times 10^{-3}$ | Pa.s |
| Air viscosity | $\mu_a$ | $1.81 \times 10^{-5}$ | Pa.s |
| Water density | $\rho_w$ | 1000 | kg/m³ |
| Air density | $\rho_a$ | 1.125 | kg/m³ |
| Surface tension | $\sigma$ | 0.072 | N/m |
| Gravitational acceleration | g | 9.81 | m/s² |
| Nusselt film thickness | $h_N = \left(\frac{3\mu^2 Re}{\rho^2 g \sin\beta}\right)^{1/3}$ | $0.1829\text{-}0.3127 \times 10^{-3}$ | m |
| Nusselt velocity | $u_N = \frac{\rho g \sin\beta h_N^2}{3\mu}$ | 0.1094 - 0.3197 | m/s |
| Film Reynolds number | $Re = \rho q / w\mu$ | 20-100 | - |
| Film Weber number | $We = \rho h_N u_N^2 / \sigma$ | 0.0304-0.4439 | - |
| Kaptiza number | $Ka = \sigma \rho^{1/3} / g^{1/3} \mu^{4/3}$ | 3363 | - |

## 5 Results and discussion

### 5.1 Natural waves

In this section, we discuss the results of the observed phenomena in the propagation of natural, unexcited waveforms on liquid film surfaces. As has been observed in previous literature [2,12,15], the propagation of waves on the film surface commences once the film Reynolds number exceeds a critical value, for the onset of primary instability, $Re_c = 5/4 \cot \beta$, triggered by the amplification of the infinitesimal disturbances at the inlet, from random noise that the system is constantly subjected to. This linearly-growing noise-caused wave grows according to the linear stability analysis of a uniform laminar flow with only a visible flat film observed before the appearance of waves becomes eminent. This primary instability is convective and hence propagated downstream with the flow and has been found to be sensitive to the noise at the substrate inlet [6,8]. However, further downstream, this small amplitude noise-driven disturbances undergo exponential growth and then become visible. They transition into two-dimensional periodic waves and later develop into solitary wave structures which are characterised by large-humped waves which are preceded by series of front running capillary waves. As these waves interact with one another, they undergo inelastic collision, with larger humped waves colliding and consuming the smaller humped waves. This wave harmonics interaction leads to the altering of the wave patterns, with the generation of three-dimensional structures on the film surface. Further downstream, the stability of the waves becomes more affected leading to the generation of highly asymmetric, three-dimensional and spatiotemporally unpredictable nonlinear waves covering the entire film surface [1, 6, 8].

In Fig. 3, we show the interface profile for the propagation of naturally developing waves on a film surface. The film Re is 100 and the substrate inclination, $\beta$ is 90˚. As can be seen, visible wave development commences with a two-dimensional periodic wave that appears on the film surface (t=0.4s). The wave slowly propagates while more capillary structures develop in its front (t=0.4 - 2.0s). Subsequently, a series of wave interaction are observed which lead to the growth of the wave hump as the waves propagate downstream. As the interaction process continues, the waves begin to yield to a three-dimensional instability (t=3.2s), which modulates the waveform transversely thereby leading to the onset of three-dimensional structures on the film. These are well captured here due to the three-dimensionality of the simulation. Essentially, curves are observed to form, which bulge downward and later expand into horseshoe structures. The legs of these structures further straighten upwards and aid to reduce the width of the flat parts. Finally, these horseshoes get detached from the flat parts and form a cluster of dimples on the film surface (t=6.24s). It must be noted that the solitary waves observed here (in contrast to those obtained when forcing is applied) are characterised by irregular spacing due to the noise amplification. As a result, they tend to yield more easily to three-dimensional instabilities, which modulates the wave profile transversely. Finally, the solitary waves observed here are also characterised by large solitary humps, preceded by series of front-running capillary waves and succeeded by a long flat tail which confirms qualitative information about naturally developing waves [5,8,10-11].

In Fig. 4, the unique features of its three-dimensionality are shown. From this, it can be seen that these disordered three-dimensional waves show interesting features of curly and bulging horseshoe structures which possess sorts of spikes at the wave hump and a succeeding flat tail that extends to the developing Nusselt film. The miniature size of the preceding capillary ripples are also well-captured with details of the effect of the transversal instability on the capillary structures also shown.

From the evolution of these waves, it can be observed that a slice in the transverse direction gives different information on the film thickness data which is quite different from that observed for a two-dimensional wave. The effect of this transversal differences is captured in Fig. 5. It can be observed that at (A), while the transversal instability is just beginning to set in, the differences in the film thickness profile at the different slice points are very minimal, with the obtained profile in Fig. 5 (A(ii)) almost the same. However, as the waves evolve into three-dimensional structures, noticeable differences are immediately seen in the slice plots [Fig. 5 (B (ii))] which becomes even more obvious in [Fig. 5 (C (ii))].

In comparison also, it can be seen that there is a general reduction in the height of the waves as the evolution continues (Fig. 5: A, B, C), which occurs as a result of the spreading out of this three-dimensional structures on the entire film surface.

Finally, the streamwise velocity plots of the waves are shown in Fig. 6. From the plot, it can be seen again that the highest streamwise velocity is obtained at the wave hump (shown in blue), while intermediate values are obtained on the flat film regions (shown in green). The least value are however observed on the capillary ripple region (shown in red). It is interesting to note that negative streamwise velocity values (depicting flow reversal) usually observed in the capillary wave region are observed here, which is in agreement with the literature [5,8-9,12,26] on the wave profile on the different regions of a solitary wave.

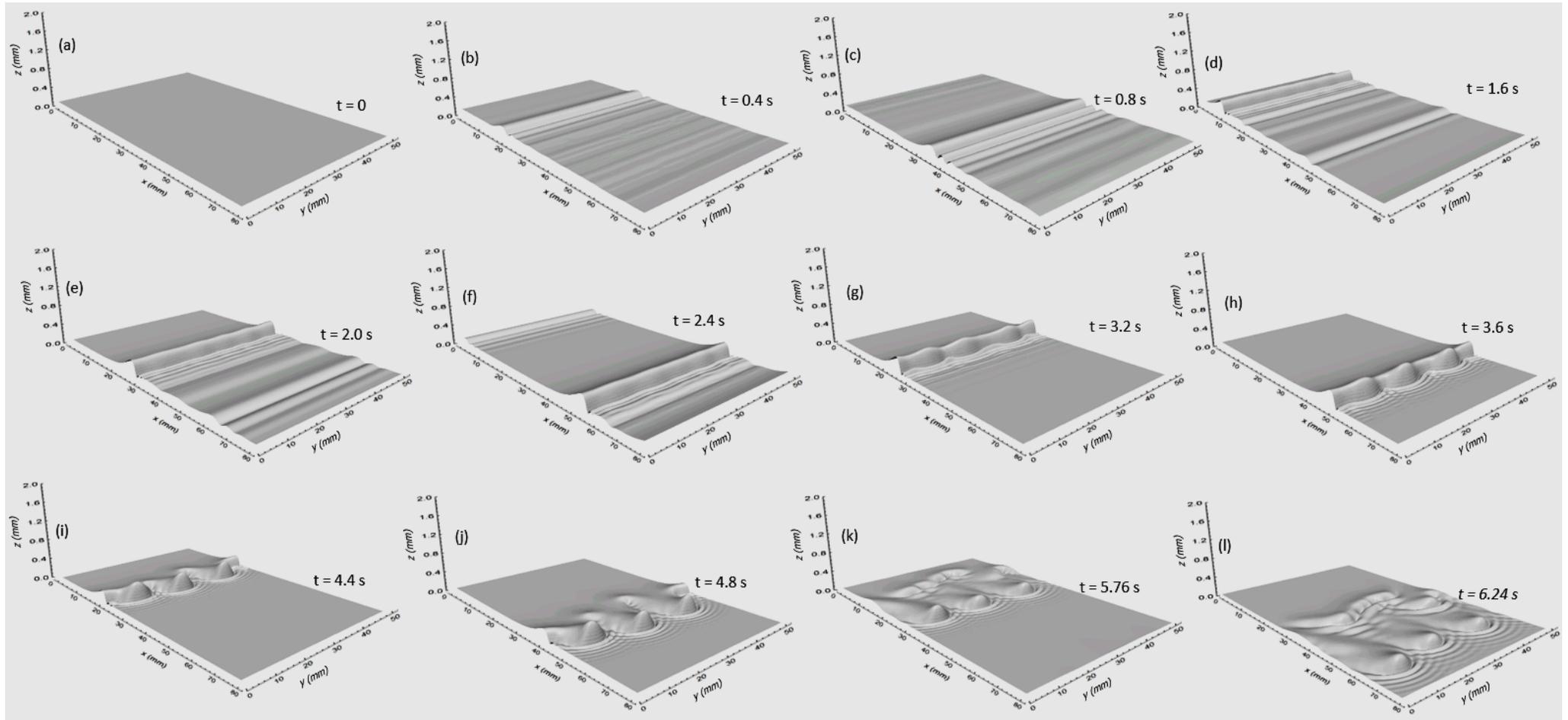

**Fig. 3** Time-propagation of naturally developing waves, in three dimensions. Simulation parameters are *Re* 100, β = 90. The interface height has been magnified 10 times to clearly reveal the wave propagation.

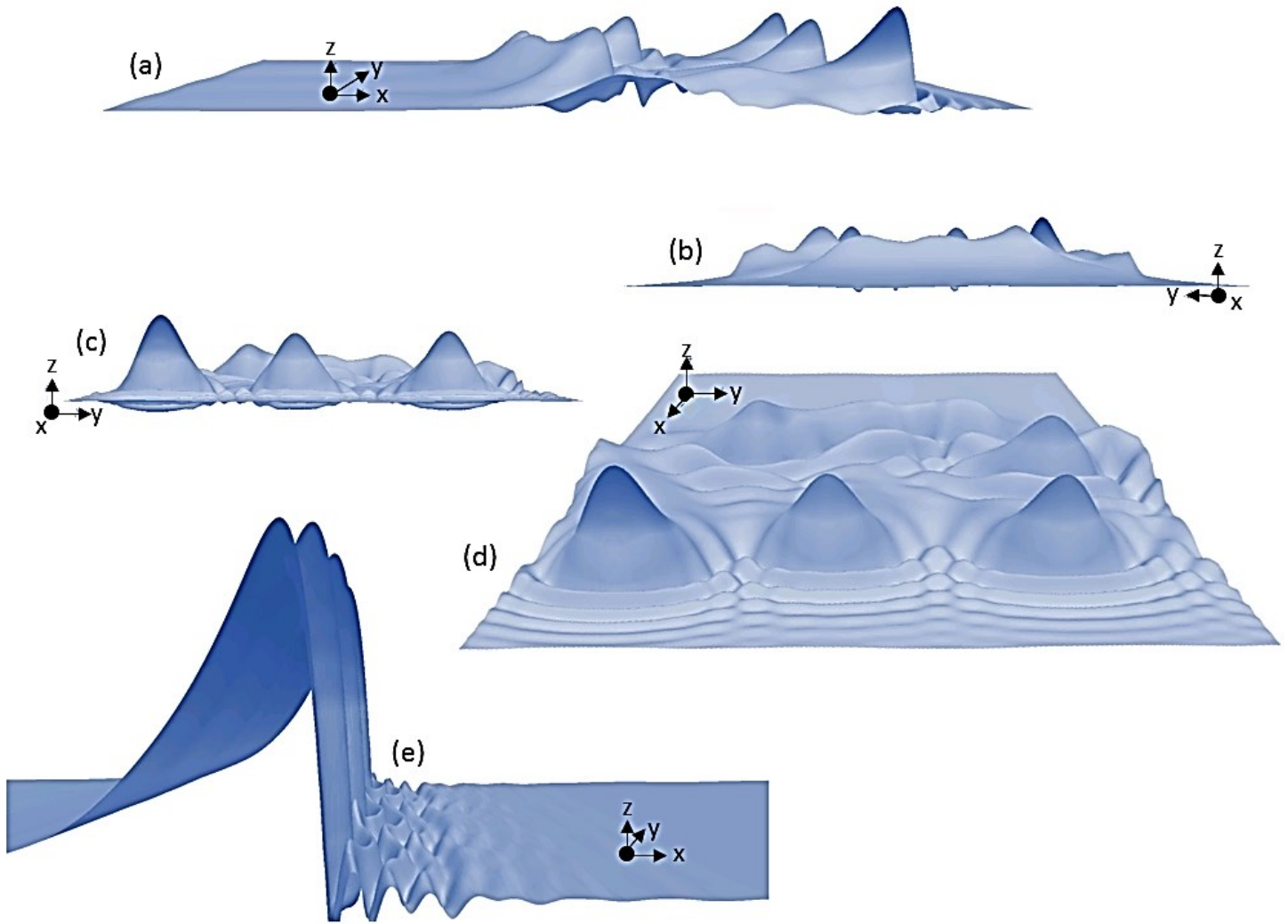

**Fig. 4** Typical interface profile of disordered three-dimensional solitary waves on a flowing liquid film. The simulation parameters are same as in Fig. 5.3. Here, the interface height has been magnified 30 times to clearly reveal minutes details of the wave features.

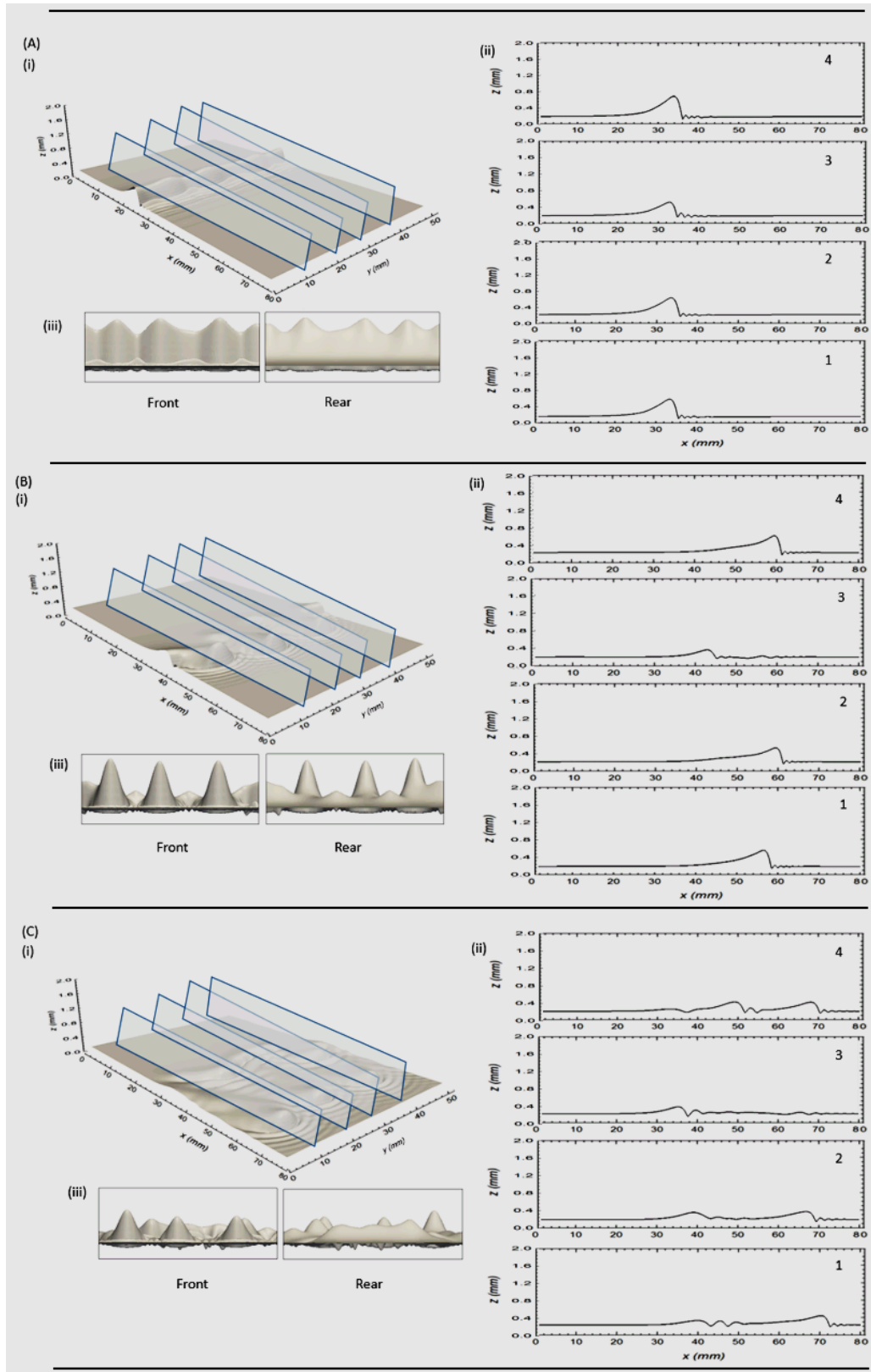

**Fig. 5** (i) Four-point transversal slice and (ii) individual film thickness plots of three-dimensional solitary waves on a naturally developing film. The slices are on the 10, 20, 30 and 40 mm points in the transverse direction and shown as plots 1, 2, 3 and 4 respectively. Close snapshot of the front and rear view of the waves are shown in (iii). Figures A, B, and C are for time 3.2 s, 4.8 s, and 6.24 s respectively. The simulation parameters are $Re$ 100, $\beta = 90$ and the interface height has been magnified 10 times to clearly reveal the wave propagation.

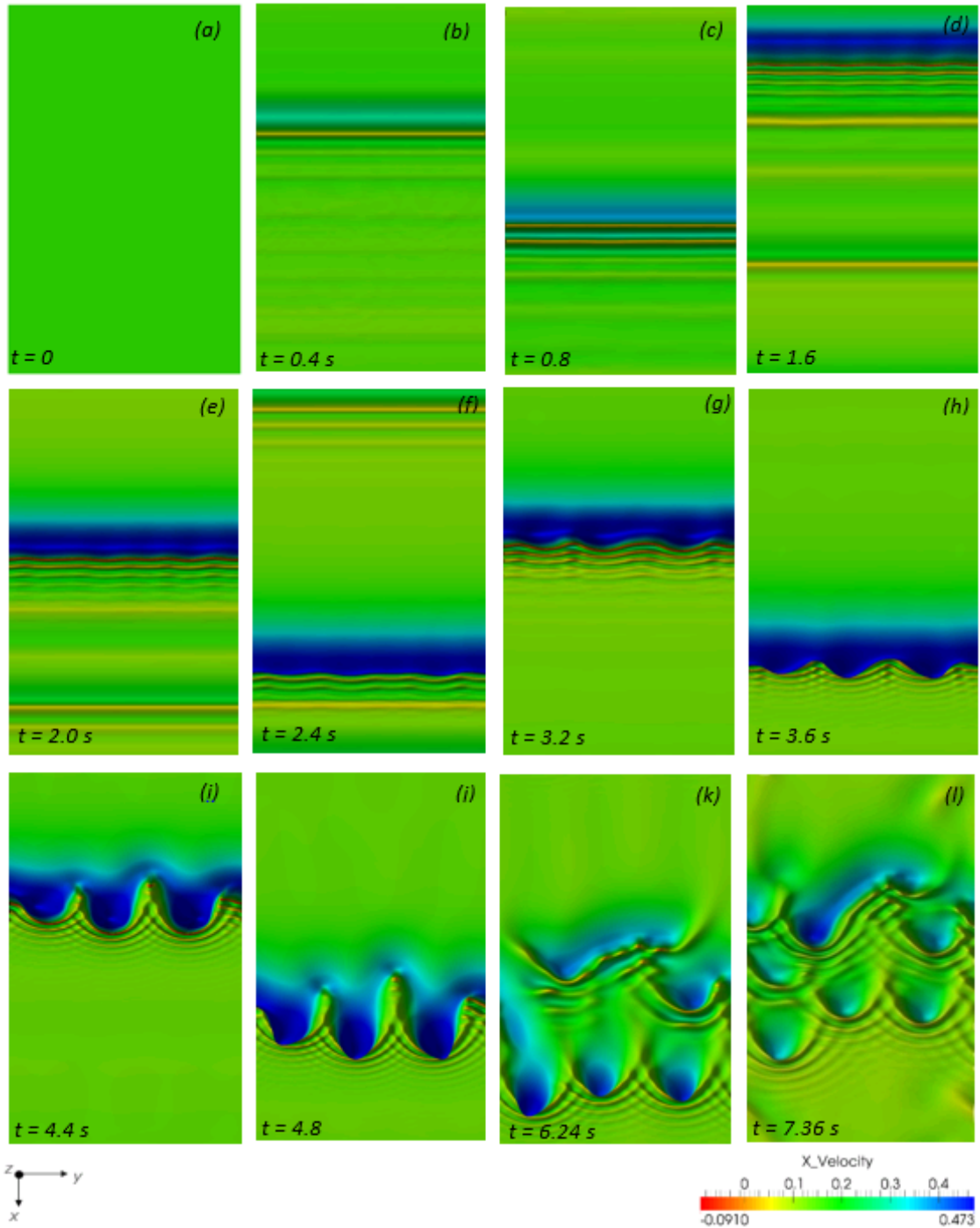

**Fig. 6** Streamwise velocity plot of naturally developing waves, in three dimensions. Simulation parameters are *Re* 100, β = 90. The interface height has been magnified 10 times to clearly reveal the wave propagation.

## 5.2 Forced waves

In this section, we consider the family of waves formed with the application of periodic inlet flow rate forcing. The need to artificially excite these waves arises due to the spatiotemporal irregularities of naturally evolving waves, hence, the application of a periodic forcing of the inlet flow rate which is able to mask the un-monochromatic noise-driven disturbance [8], thereby controlling the wave evolution dynamics [6,8-9,11].

**5.2.1 Gamma I waves:** These family of waves are known to be characterised by wide and closely-joined peaks which appear sinusoidal and travel a distance of several wavelengths downstream appearing like quasi-stationary waves. They are obtained when high forcing frequency is applied [9,24-25]. In their evolution, finite-amplitude waves are observed immediately after the saturation of the linearly growing waves. During their propagation, an uneven separation between these waves is usually noticed which occurs due to the onset of secondary instabilities [1,9]. When get subjected to these secondary instabilities, they transition into regular three-dimensional patterns [2,9,-11] which has been detailed in the literature.

The interface profile as well as information on the time-propagation and film thickness of the gamma I wave family obtained in this work is shown in Fig. 7. As can be seen from the plot, waves appear on the film surface from an undisturbed Nusselt flat film. The initial two-dimensional wave propagates downstream with the film flow, undergoing complex interactions with subsidiary fronts. Soon, multiple humps are seen on the film surface, with peaks that grow closer with time. After a period of propagation, the wave shape typically becomes alike with the peaks closely knit together and travelling at a similar speed and also sharing the same hump height. Due to the short streamwise distance travelled by the wave, there is no spanwise deformation, hence the wave shape appears similar to that of a two-dimensional simulation. In Fig. 8, the streamwise velocity plots of the waves are given. From the velocity plot, again, it can be seen that the velocity at the wave hump is the highest (shown in blue), while a reduced value is obtained just before the wave trough (shown in green). The least streamwise velocity is however obtained at this trough (shown in red), which is in agreement with previous findings in the literature on wave velocity profile in different regions of the waveform [6,8-9,11].

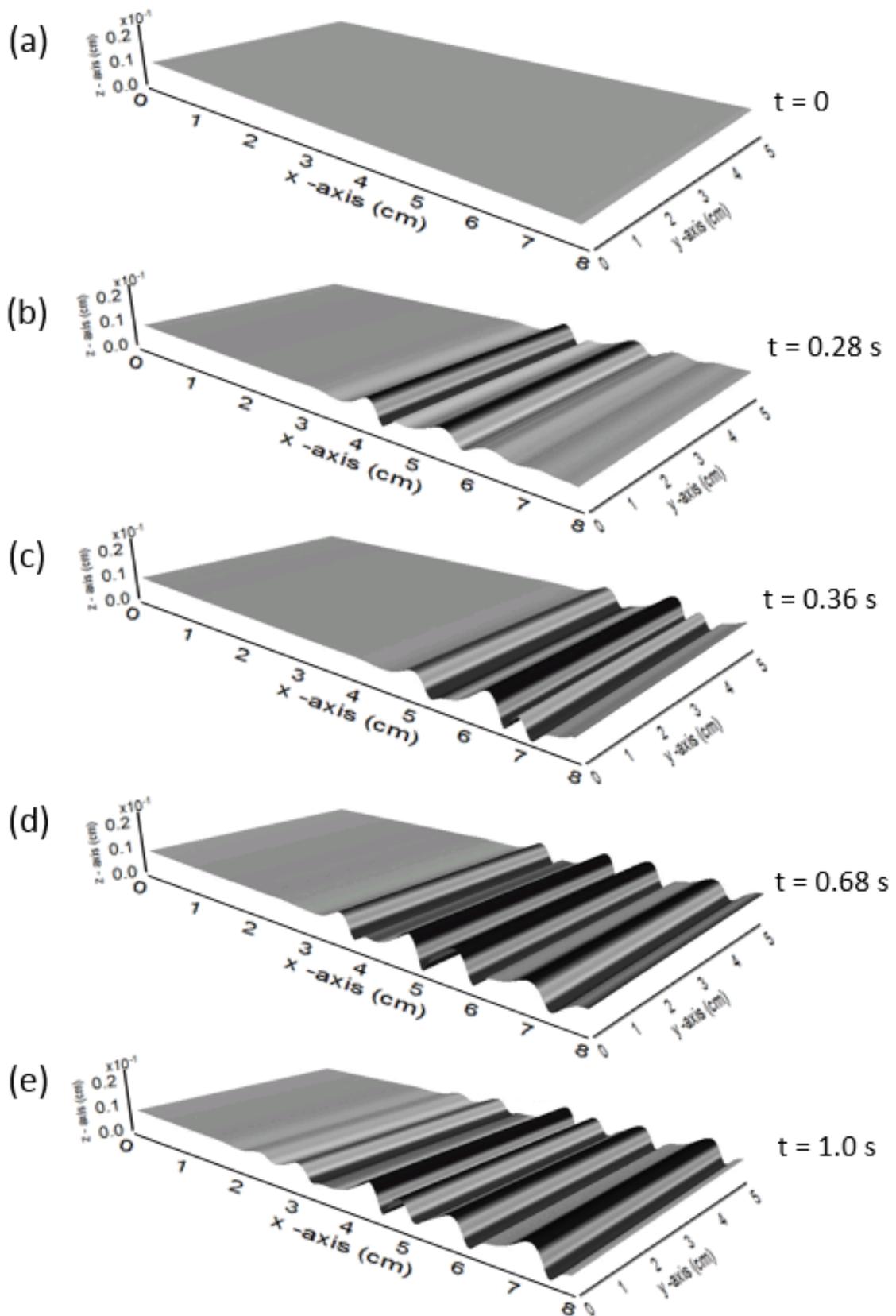

**Fig. 7** Interface profile for the propagation of gamma I wave family in three dimensions (code Blue). Simulation parameters are *Re* 20, β = 90, F =27 Hz. The interface height has been magnified to clearly reveal the wave propagation. The computed wave speed is 0.213 m/s

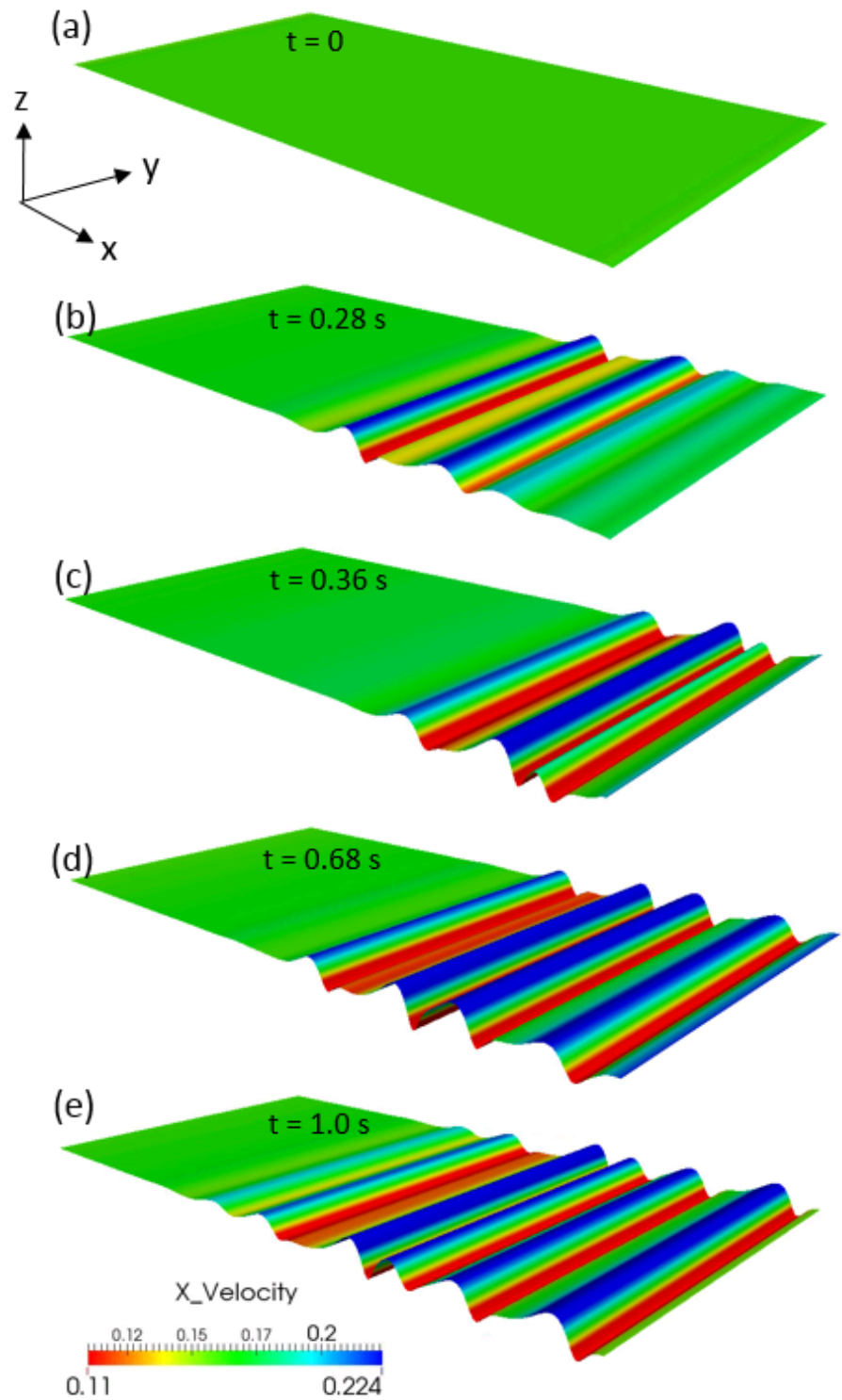

**Fig. 8** Streamwise velocity plot of the gamma I wave family. Simulation parameters are same as in Fig. 5.24. The interface height has been magnified to clearly reveal the wave propagation.

**5.2.2 Gamma II waves:** In the development of this wave family, the initial wave separates and develops steep fronts and stretched tails as the wave moves downstream. The subsidiary wavefronts nucleate while primary peaks grow larger and later saturate. Soon after the saturation of this linearly growing waves, solitary waves appear, which travel further downstream until they reach a stationary state in which successive pulses are nearly identical [9,11-13]. The waves are first seen as two-dimensional structures (with no spanwise dependence or velocity component: the wave amplitude varying only in the streamwise direction). However, further downstream, they yield to secondary instabilities that mainly affects their hump and soon transit into three-dimensional solitary waves [9-10,15]. They are known to be fast and also predominate for long waves and are characterised by tall and widely-separated narrow peaks, preceded by a series of front running capillary waves or ripples [8,10,14]. They are also usually obtained at low excitation frequencies and exhibit inelastic collision as higher amplitude - faster-travelling waves consume the smaller pulses [8,24].

The interface profile of the gamma II wave family obtained in this work is shown in Fig. 9. Here, it can be seen that wave growth commences just as in the gamma I wave family, from the initially flat film (Fig. 9a), but here, dissimilar hump heights are quickly observed (Fig. 9b). Larger humped waves are also observed to collide with smaller humped waves (Fig. 9c) in an inelastic manner leading to yet larger humped structures (Fig. 9d-e).

Due to this, the waves travel faster. Essential features of these waves include large solitary humps, which are preceded by a series of front-running capillary waves and succeeded by a long flat tail. Here also, no spanwise deformations are significantly observed due to the short streamwise distance travelled by these waves. Essentially, this behaviour follows closely the basic features of the gamma II waves as detailed in the literature [8,10,24-25]. In Fig. 10, the streamwise velocity plot of this wave family is further provided. Again, it can be seen that the velocity at the hump regions are the highest (shown as blue), while intermediate values are observed just before the wave trough and on the flat tail succeeding the humps (shown as green). The least streamwise velocity values, however, are seen in the capillary wave region (shown as red). Finally, it is interesting to note that the peak velocity observed for this waveform is greater than that observed for the gamma I wave family. This is in precise agreement with data in the literature [1-3, 14-15], that gamma II waves travel faster than their counterpart (gamma I), essentially, due to the inelastic collision that they (gamma II waves) usually go through, leading to the development of large-humped solitary waves, which are able to travel faster on the film surface.

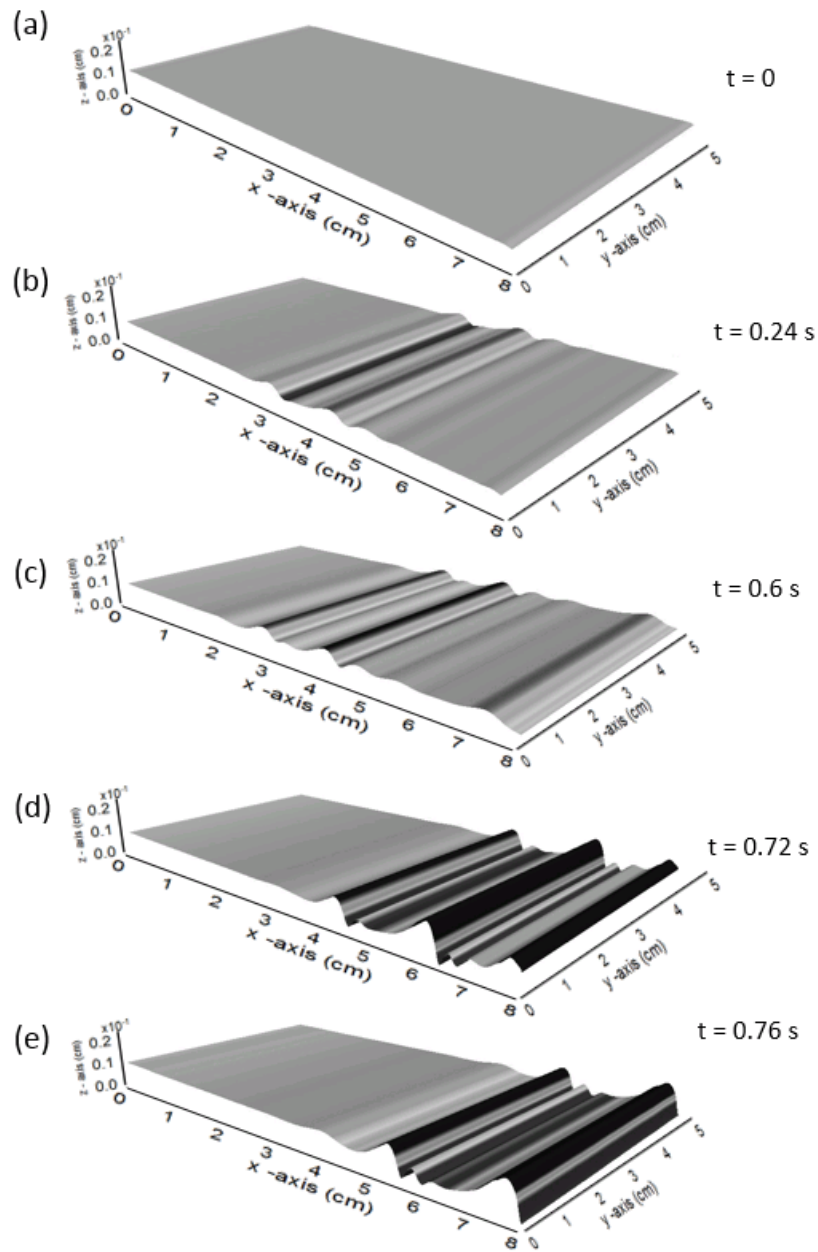

**Fig. 9** Interface profile for the propagation of gamma II wave family in three dimensions (code Blue). Simulation parameters are $Re$ 62, $\beta = 90$, F =15 Hz. The interface height has been magnified to clearly reveal the wave propagation. The computed wave speed is 0.233 m/s

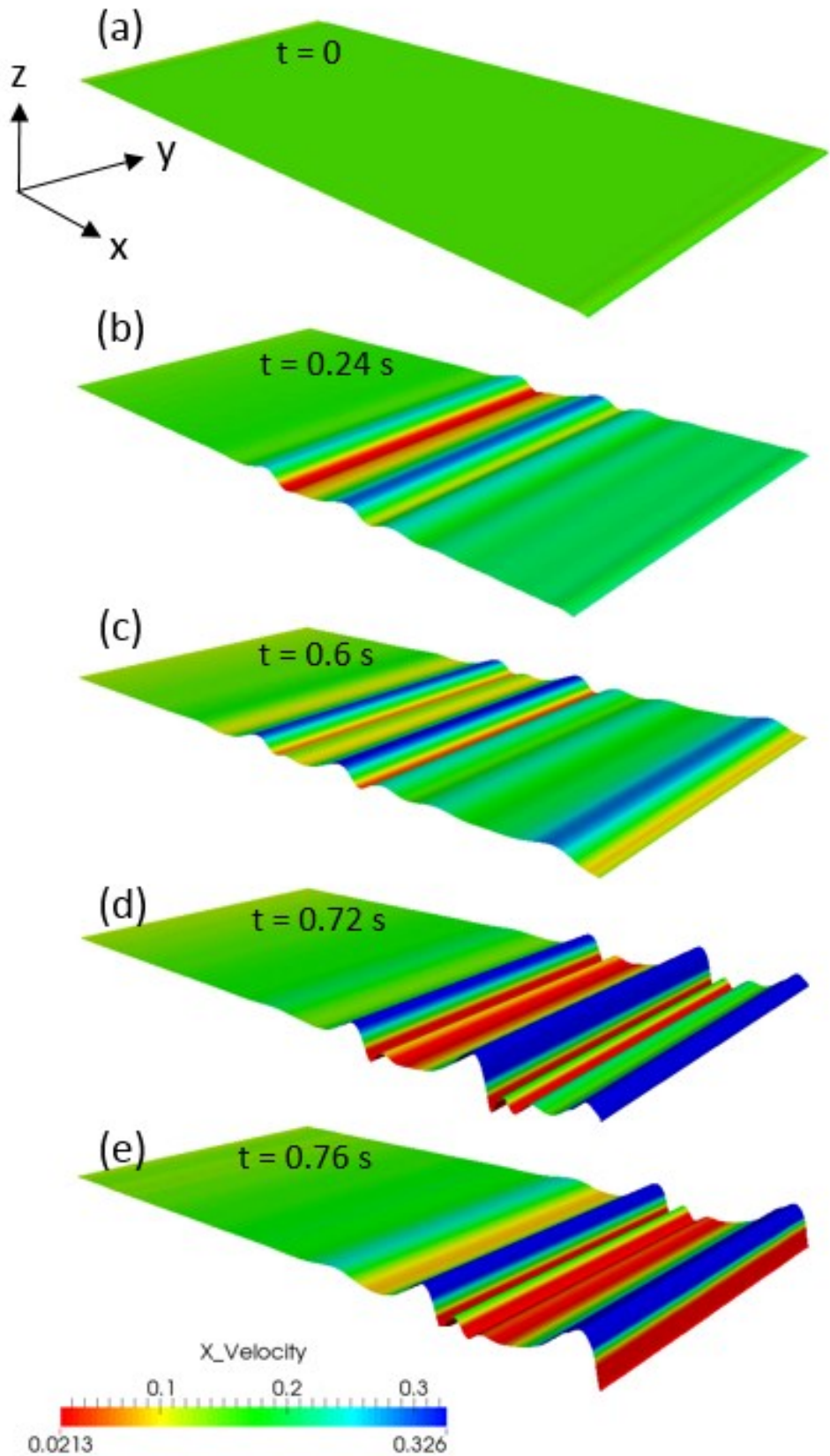

**Fig. 10** Streamwise velocity plot of the gamma II wave family in three dimensions (code Blue). Simulation parameters are *Re* 62, $\beta = 90$, F =15 Hz. The interface height has been magnified to clearly reveal the wave propagation. The computed wave speed is 0.233 m/s

### 5.3 Comparison with experimental results

In this section, we further point out a number of qualitative features exhibited by the solitary waves obtained in this study and compare them with the literature. It is important to note first that the solitary waves obtained were fully developed and characterised by a series of front running capillary waves preceding a large humped wave. Spanwise deformations of two-dimensional waveforms leading to the onset of three-dimensional horseshoes structures; inelastic collision leading to the creation of larger humped waves, as well as spatio-temporal irregularities of wave dynamics on the film surface were all observed. The wave velocity obtained i.e. 0.213 m/s also compare well with the experimental data of Kapitza and Kapitza [7] who found a velocity of 0.217 m/s for *Re* 20, $\beta = 90$, F =27 Hz.

## 6 Conclusion

In this paper, a full Direct Numerical Simulation of three-dimensional falling liquid films is presented. An hybrid front-tracking/level-set hybrid method is employed which has distinct advantages of accurate capture of the details of the interface topology while outstripping issues with mass conservation, numerical diffusion, or spurious interface profiles. Three categories of simulations were carried out. First, simulations involving a naturally developing film were considered. An "infinite" domain was specified through the imposition of periodic boundary conditions at the film outlet. Obtained results show vivid transitions undergone by the waveforms from the two-dimensional state to three-dimensional disordered structures, possessing large solitary humps, preceded by series of front-running capillary waves and succeeded by a long flat tail. Features which two-dimensional simulations leave only to the imagination were clearly portrayed and found in agreement with previous predictions from the literature. Three-dimensional simulations of artificially forced waves were then carried out which gave rise to two different wave families viz. the gamma I and II waveforms. Important parameters of these waveforms i.e. speed, wavelength and peak height were then compared with the existing two-dimensional benchmark cases in the literature and excellent agreement found. In all, the three-dimensionality of the simulation ensured that features such as the transversal instabilities; horseshoe structures formation, detachment, formation of film surface "dimples" were clearly portrayed. With inlet-forcing, however, regular wave spacing were observed in contrast to those on natural and unforced films which are characterised by irregular spacing due to the noise amplification.

# Acknowledgement

The project was sponsored by the Petroleum Technology Development Fund (PTDF, Nigeria). I.A. also appreciates Drs. Damir Juric, Jalel Chergui and Sheughon who originally developed the numerical code.